\documentclass[a4paper, amsfonts, amssymb, amsmath, reprint, showkeys, nofootinbib, twoside]{revtex4-1}
\usepackage[english]{babel}
\pdfoutput=1
\usepackage[utf8]{inputenc}
\usepackage[colorinlistoftodos, color=green!40, prependcaption]{todonotes}
\usepackage{amsthm}
\usepackage{mathtools}
\usepackage{physics}
\usepackage{xcolor}
\usepackage{graphicx}
\usepackage[left=23mm,right=13mm,top=35mm,columnsep=15pt]{geometry} 
\usepackage{adjustbox}
\usepackage{placeins}
\usepackage[T1]{fontenc}
\usepackage{lipsum}
\usepackage{csquotes}
\usepackage[pdftex, pdftitle={Article}, pdfauthor={Author}]{hyperref}% For hyperlinks in the PDF
\hypersetup{
    colorlinks=true,
    linkcolor=blue,
    filecolor=magenta,      
    urlcolor=cyan,
}
\begin{document}
\title{Active binary mixtures of fast and slow hard spheres}

\author{Thomas Kolb}
    %\email[Correspondence email address: ]{email@institution.com}% Your name
    \affiliation{Department of Chemistry, University of North Carolina at Chapel Hill, Chapel Hill, NC, USA}
\author{Daphne Klotsa}
    \email[Correspondence email address: ]{dklotsa@email.unc.edu}% Your name
    \affiliation{Department of Applied Physical Sciences, University of North Carolina at Chapel Hill, Chapel Hill, NC, USA}

\date{\today} % Leave empty to omit a date

\begin{abstract}
We computationally studied the phase behavior and dynamics of binary mixtures of active particles, where each \lq species\rq \ had distinct activities leading to distinct velocities, fast and slow.
We obtained phase diagrams demonstrating motility-induced phase separation (MIPS) upon varying the activity and concentration of each species, and extended current kinetic theory of active/passive mixtures to active/active mixtures. We discovered two regimes of behavior quantified through the participation of each species in the dense phase compared to their monodisperse counterparts. 
In regime I (active/passive and active/weakly-active), we found that the dense phase was segregated by particle type into domains of fast and slow particles. Moreover, fast particles were suppressed from entering the dense phase while slow particles were enhanced entering the dense phase, compared to monodisperse systems of all-fast or all-slow particles. These effects decayed asymptotically as the activity of the slow species increased, approaching the activity of the fast species until they were negligible (regime II). In regime II, the dense phase was homogeneously mixed and each species participated in the dense phase as if it were it a monodisperse system; each species behaved as if it weren't mixed at all.
Finally, we collapsed our data defining two quantities that measure the total activity of the system. We thus found expressions that predict, \textit{a priori}, the percentage of each particle type that participates in the dense phase through MIPS.
\end{abstract}

\keywords{active matter, soft matter, self-assembly}

\maketitle

%%%MAIN TEXT%%%%
\section{Introduction} \label{intro}

From schools of fish to swarms of bacteria, the complex, collective behavior found in natural swarms and flocks has spurred an interest in systems where constituents locally convert energy into motion, referred to as active matter systems. At high constituent particle densities, such systems can be viewed as a ``living'' material with the ability to do work and adapt to stimuli, heal itself, \textit{etc.} similar to \textit{e.g.} biological tissue. The remarkable properties of active matter (self-healing, responsiveness, adaption, \textit{etc.}) can be leveraged in experimental setups to complete tasks and do work on the microscopic scale, \textit{e.g.} bacteria can be used to turn a microscopic gear~\cite{DiLeonardo2010}.

Simple models have been developed to capture the emergent behavior of active matter, including the Vicsek model~\cite{Vicsek1995f} inspired by bird flocking, run-and-tumble~\cite{BERG1972,Schnitzer1993,Cates2015a} by bacterial swarming, and the active Brownian particle (ABP)~\cite{Schimansky-Geier1995,Cates2015a} by active self-propelled colloids. 
These models demonstrate a variety of unusual nonequilibrium states, such as robust flocking bands~\cite{Solon2015c}, upstream swimming of bacteria~\cite{Nash2010} and a negative surface tension for active droplets~\cite{Patch2018} respectively. Since the inception of this field, many attempts have been made to map these systems (and the counter-intuitive steady-state behaviors they present) on to equilibrium systems~\cite{Rein2016,Farage2015EffectiveSuspensions,Marconi2016VelocitySystems, Fily2017}. However, the mapping of any pair-wise potential on to Brownian colloids to recreate non-equilibrium systems has not succeeded thus far. As such, simulation continues to be a good method of investigating active matter.

The standard ABP model consists of a monodisperse system of active hard spheres confined to a plane. In the absence of any attractive potential between particles, and at sufficient activity, the system phase separated (either gas/liquid or gas/solid), in a process known as motility-induced phase separation (MIPS)~\cite{Redner2013a,Cates2015a}. The predicted MIPS was recapitulated in experiment with light-activated "colloidal surfers"~\cite{Palacci2013c} and has since been adapted to a variety of experimental set-ups which depend on either light, magnetic/electric fields, or acoustics to induce phoresis~\cite{Driscoll2019}. 
The ABP model has also been implemented with additional complexities \textit{e.g.} confinement~\cite{Yang2014a, Fily2017, Das2018, Mahapatra2017}, anisotropy in particle shape~\cite{Kurzthaler2016}, particle interactions~\cite{Pu2017} and size polydispersity~\cite{Yang2014a, Fily2014, Dolai2018} leading to the discovery of new emergent phenomena, such as particle sorting and segregation.

Recent work has shown that mixtures of active and passive Brownian particles also display MIPS, similar to a monodisperse active system~\cite{Stenhammar2015, Wittkowski2017}. Under certain parameters, for example, steady states emerge wherein particles partially segregate into active and passive domains~\cite{Stenhammar2015, Wittkowski2017}. 
Mixtures provide an additional degree of control for performing specific functions. For example, doping a monodisperse Brownian system with active particles can help anneal crystals and rapidly relax jammed states~\cite{Ramananarivo2019, Omar2019}. Many active living systems interact not only with passive Brownian particles and obstacles but also with other active species, \textit{e.g.} various walking speeds of people in a crowd, sheep being herded by sheepdogs, or the orientationally-dependent swim speed of \textit{Pseudomonas putida}~\cite{Theves2013}. In fact, biology provides a wealth of relevant active/active systems, ranging from the micro~\cite{Cates2012a, Ben-Jacob2016, Pessot2019} to macro-scale~\cite{Sengupta2011}. Simple physics models like the ABP promote a deeper understanding of the function of activity in these systems.
However, the ABP model has not been implemented in the computational study of active/active systems. Instead, other means of activity have been examined: mixtures of active rotors\cite{Nguyen2014, Yeo2016}, particles propelled by distinct colored noise\cite{Wittmann2018}, polymers and colloids equilibrating with distinct heat baths~\cite{weber2016,Smrek2017,grosberg2015} or in systems of monodisperse ABPs with moving obstacles\cite{Ahmed2017}.
While there is a lot of work on monodisperse ABPs and more recently on mixtures of active and passive particles, the ABP model has not yet been applied to mixtures of active particles with distinct, non-zero, activity. 

In this paper, we computationally studied binary active mixtures of `fast' and `slow' particles, using the ABP model. We first obtained phase diagrams for active/active mixtures relating area fraction and activity, and extended the kinetic theory by Stenhamar \textit{et al.}~\cite{Stenhammar2015} to include two active species. We discovered two regimes of behavior quantified through the participation of each species (fast and slow) in the dense phase compared to a monodisperse system of each type. 
In regime I (active/passive and active/weakly-active), we found that the MIPS phase was strongly segregated by particle type; the edge of the dense phase was comprised primarily of fast particles, with domains of slow particles populating the cluster interior. Additionally, in the first regime, fast particles were suppressed from entering the dense phase while slow particles were enhanced entering the dense phase, compared to a monodisperse system of all-fast or all-slow particles. These effects decayed asymptotically until they were negligible in regime II. In regime II, the dense phase was homogeneously mixed and each species participated in the dense phase as it would were it a monodisperse system; each species in the system behaves as if it weren't mixed at all.
Finally, we defined the net activity, which is an average of each constituent particle's activity weighted by its area fraction, and the ratio net activity which takes into account also the relative ratio between fast and slow activities. Through collapse of our data with these two quantities, we found expressions that predict, \textit{a priori}, the percentage of each particle type that will participate in the dense and gas phases through MIPS.

The structure of this paper is as follows. In section~\ref{methods} we outline the theoretical and computational methods, and simulation details for the active systems studied. In section~\ref{background}, we discuss the background that relates to our work, \textit{i.e.} previous studies of monodisperse active and mixtures of active/passive systems. We describe our results in section~\ref{results}, and end with conclusions and outlook in section~\ref{conclusion}. 

\section{Methods} \label{methods}

Our system approximates the physics of active colloids confined to a plane. We used $N$ spheres, each with a body axis that indicates a  particle's direction of self-propulsion (restricted to be in-plane). Particles translate and rotate according to the overdamped Langevin equations of motion for translation and rotation:
\begin{equation}\label{brownmotion}
\gamma\dot{\vec{r_{i}}}=F_{LJ}(\vec{r_{i}})+F_{act}\widehat{p_{i}}+\gamma\sqrt{2D_{t}}\Lambda_{i}^{t}
\end{equation}
\begin{equation}
\dot{\theta_{i}}=\sqrt{2D_{r}}\Lambda_{i}^{r} \quad,
\end{equation}
where $r_{i}$ is the position of particle $i$, $\widehat{p_{i}}=(\cos{\theta_{i}},\sin{\theta_{i}})$ gives its orientation in the $xy$-plane, $F_{LJ}$ is the conservative force due to pairwise interactions, $F_{act}$ is a particle's active force, $D_{t}$ and $D_{r}$ are the translational and rotational diffusion constants, and $\gamma$ is the drag coefficient. The random force incorporates unit variance Gaussian white noise, implemented through $\Lambda^{t}$ for translation and $\Lambda^{r}$ for rotation such that $\langle \Lambda_{i} \rangle=0$ and $\langle \Lambda_{i}(t)\Lambda_{j}(t')\rangle=\delta_{ij}\delta(t-t')$. In overdamped systems (low Reynolds number), the Stokes-Einstein equation gives $D_{t}=\frac{k_{B}T}{3\pi\eta\sigma}=\frac{k_{B}T}{\gamma}$ and $D_{r}=\frac{3D_{t}}{\sigma^2}$, where $\sigma$ is the particle diameter, $\eta$ the dynamic viscosity, and $k_{B}T$ the thermal energy. We used $\sigma$ as the non-dimensional unit of distance. 

Activity was quantified through the P\'{e}clet number:
\begin{equation}
Pe=\frac{3v_{0}\tau_{r}}{\sigma},
\end{equation}
where $(v_{0})$ is the intrinsic swim speed, \textit{i.e.} the speed an active particle has under the action of the active force $F_{act}$ (in the absence of collisions), and $\tau_{r}$ is the reorientation time, given by $\tau_{r}=D_{r}^{-1}$. The corresponding persistence length is $l_{p}=v_{0}\tau_{r}$ and quantifies how far an active particle travels maintaining its direction before it undergoes thermal rotation. 

In simulations, we varied the active force, $F_{act}=v_{0}\gamma\widehat{p}$ via the intrinsic swim speed $v_{0}$. This was intentionally distinct from other works, where activity was modulated through the diffusion constant by varying the system temperature~\cite{Stenhammar2014, Stenhammar2015, Wittkowski2017}. We chose to modulate activity via the swim speed for the following reason. In our binary system, there are two active forces of different magnitude, one for each species, giving two intrinsic speeds: fast and slow. Inevitably, with a temperature-variant model, fast and slow particles would have two different temperatures in the same system - thus producing two different random forces. The differences between velocity and temperature-variant models become clearer if we examine the motion of particles on a set timescale. In our velocity-variant model a fast particle will experience a greater active force and move a greater distance than a slow particle. In a temperature-variant model, fast and slow particles will move with the same velocity and thereby experience the same active force. Thus in the temperature-variant model particle types differ in how long they maintain their direction; so, the rotational diffusion for slow particles is larger than that of the fast particles.
Note that, in examinations of MIPS, it has been shown that particle rotation controls the rate of desorption from the dense phase, such that the faster the particle rotation the larger the rate of desorption from the dense phase~\cite{Redner2013a, Redner2013}. As a temperature-variant model directly changes the rotational diffusion between particle types, it would also bias the composition of the cluster edges to have more of the slowly-rotating, fast species. 

All particles, regardless of type, experience excluded volume interactions via a cutoff $(r_{cut}=\sqrt[\leftroot{-3}\uproot{3}6]{2}\sigma)$ and shifted Lennard-Jones potential,
\begin{equation}\label{ljPotential}
U({r_{i,j}}) = 
	\begin{cases} 
      4\epsilon [(\frac{\sigma}{{r_{i,j}}})^{12} - (\frac{\sigma}{{r_{i,j}}})^6]+\epsilon & 0 \leq {r_{i,j}}\leq \sqrt[\leftroot{-3}\uproot{3}6]{2}\sigma \\
      0 & {r_{i,j}}> \sqrt[\leftroot{-3}\uproot{3}6]{2}\sigma, \\
   \end{cases}
\end{equation}
with corresponding force,
\begin{equation}\label{ljForce}
F({r_{i,j}\le r_{cut}}) = 24\epsilon \Bigg(2\frac{\sigma^{12}}{r^{13}} - \frac{\sigma^{6}}{r^{7}}\Bigg)
\end{equation}
where $r_{ij}$ is the interparticle distance between the $i^{th}$ and $j^{th}$ particles. To nondimensionalize our system we set $\sigma=1$ as the particle diameter, and $\epsilon=\epsilon_{act}+\epsilon_{Brownian}=k_{B}T(\alpha + 10)$ as the potential well depth, where $\alpha$ is an activity-dependent coefficient ($\alpha \sim F_{act}\sigma$) implemented to maintain a constant, hard sphere, particle diameter. Note that, for passive particles, the repulsive well depth reduces to $\epsilon=10k_{B}T$.
To maintain a constant particle diameter we consider the greatest force present at a given set of parameters; in a Brownian system this is the thermal force. As first acknowledged by Stenhammar \textit{et al.}~\cite{Stenhammar2014}, this is not the case in active systems because the active force typically exceeds the thermal force. If we used the thermal force to set the repulsive strength, particles would experience a degree of overlap (up to $\sigma_{effective}=0.75\sigma_{intended}$ for the activities studied here).
Stenhammar \textit{et al.}~\cite{Stenhammar2014} suggested setting the potential well depth with coefficient $\alpha=\frac{F_{act}\sigma}{24}$ (derived by setting $\sigma=r=1$, and $F_{LJ}=F_{act}$ in eqn.~\ref{ljForce}).
To test the validity of this ratio in our system we performed simulations on monodisperse active systems as well as binary active/passive, and active/active systems, with $\epsilon=k_{B}T$ at total area fraction, $\phi=0.6$. We then computed the center-to-center particle distance ($\delta$) and plotted a histogram for $\delta\le r_{cut}$ on the simulation data.

For a hard sphere system, we expect a narrow distribution of emergent particle diameters with mode center-to-center distance equal to the intended diameter $(\delta_{mode}=\sigma_{intended}=1)$.

\begin{figure}[!htb]
\center{\includegraphics[height=4.cm,trim={0 0 0 0}, clip]
{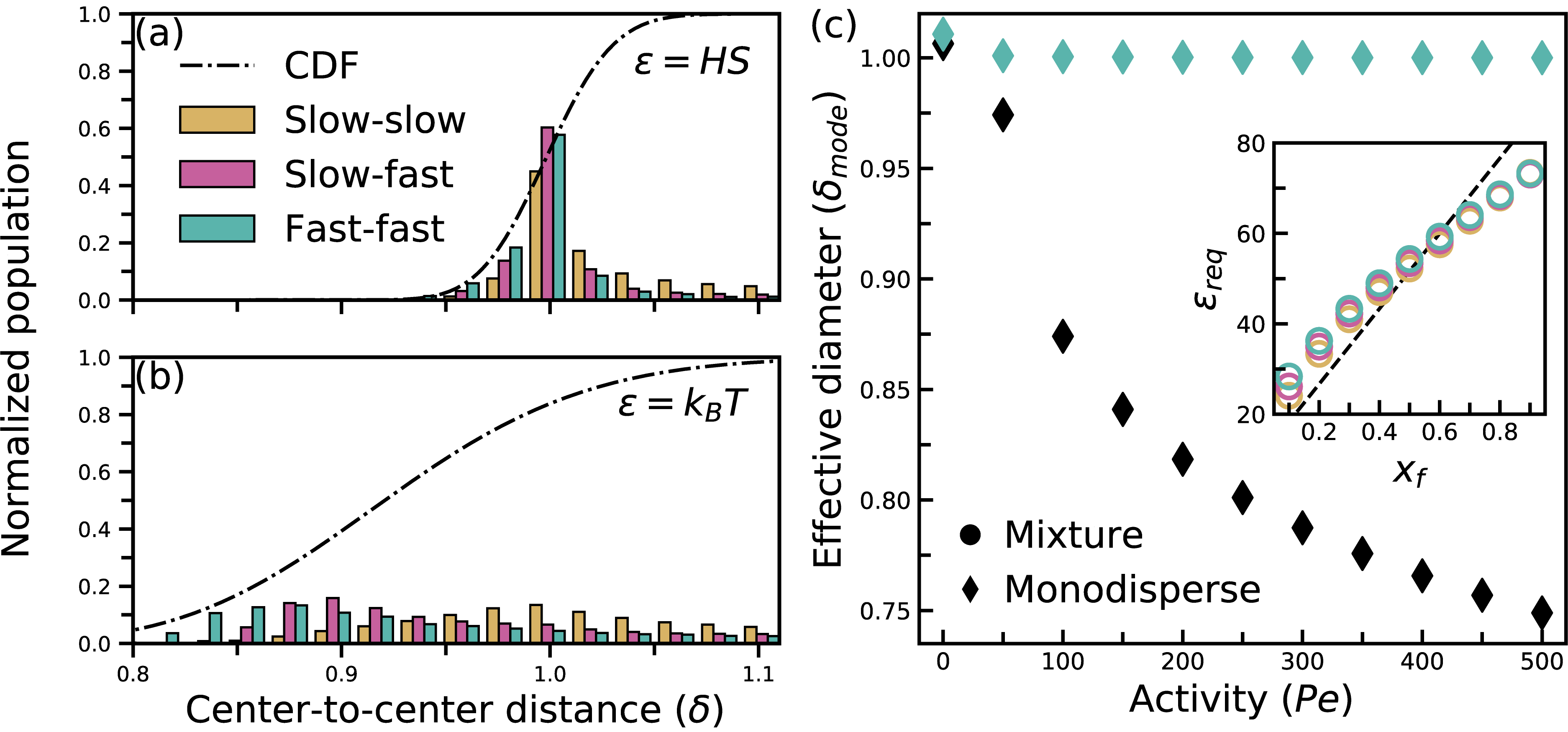}}
\caption{\label{potential_fig}
Comparison of particle activity and interparticle distance for particle pairs within the Lennard-Jones interaction cutoff $(0 \le r \le r_{cut})$. 
(Left) Histograms of the normalized center-to-center distance showing each interaction type as well as the cumulative distribution function (CDF) overlaid. CDF displays probability of a particle having diameter $\delta \le x$.
(a) Histogram and CDF for simulations performed using $\epsilon$ from equation~\ref{netAlpha} demonstrates that no particles have center-to-center distance $\delta \le 0.95$ while $50\%$ of particles have center-to-center distance $0.95 < \delta \le 1.00$. Data are computed from an active/passive simulation with $\epsilon=k_{B}T$, $x_{f}=0.1$ and $Pe_{f}=500$.
(b) Histogram and CDF for simulations using $\epsilon=k_{B}T$ where all types of interaction data deviate well below the intended diameter. (c) Plot of the effective diameter measured in simulation as a function of particle activity. Simulation data for monodisperse simulations with well depth $\epsilon=k_{B}T$ (black) shows drastic particle overlap as activity is increased. Data for systems simulated with interaction potential according to equation~\ref{netAlpha} (teal), maintains a constant mode center-to-center distance that is equal to the intended diameter ($=1$ for hard spheres). Inset shows predicted $\epsilon$ both from equation~\ref{netAlpha} (dashed line) and computed from emergent simulation data with respect to system composition. Required well depth is computed for slow-slow (gold), slow-fast (pink), and fast-fast (teal) interactions.
}
\end{figure}

Center-to-center distances less than the intended diameter $(\delta<1)$
indicate that there are instances when the forces are larger than accounted for by the repulisve pairwise potential and thus particles become slightly ``soft''. Indeed, this commonly occurs when the repulsive depth is set by the thermal force $(\epsilon=k_{B}T)$. The center-to-center distance histogram for these data exhibited values far lower than the intended particle diameter (fig.~\ref{potential_fig}b, $\delta<1$). 

To prevent this, we extracted the mode of the distribution ($\delta_{mode}$) at several activities and particle fractions and substituted back into equation~\ref{ljForce} as $r$, to obtain the mode force experienced by the particles.
We then used that force on the left-hand side of equation~\ref{ljForce}, with $\sigma=r_{i,j}=1$ (to have hard spheres) and solved for the interaction well depth required to prevent particle overlap $(\epsilon_{required})$. For monodisperse systems, plotting the effective diameter as a function of activity produced an interaction well depth very similar to that predicted by ref.~\citenum{Stenhammar2014}: $\epsilon_{req}=\frac{4F_{act}\sigma}{24}$ (fig.~\ref{potential_fig}.c, black diamonds), where an increasing active force corresponds to a smaller effective diameter and therefore a larger required repulsive force (to correct for emergent particle softness).

In a binary mixture three interaction potentials must be considered: fast-fast, fast-slow, and slow-slow. As such, we broke down our center-to-center analysis by interaction type. Our data showed that all interactions for a given simulation predict the same repulsive well depth ($\epsilon_{req}$, fig.~\ref{potential_fig}c inset). Therefore, we can implement equal values of $\epsilon$ for all interaction types. Additionally, we found that the magnitude of the potential well depth should be weighted according to the particle fraction of each type (fig.~\ref{potential_fig}.c inset). This gave the following relationship for the interaction well depth that was used in our studies,

\begin{equation}\label{netAlpha}
    \epsilon_{req.}=\left(\frac{4(F_{s}x_{s}+F_{f}x_{f})}{24} + 10\right)k_{B}T
\end{equation}

\noindent As the repulsive well depth incorporates the fraction of each species, this implementation reproduces the relation that was determined for monodisperse active systems as well ($\alpha=\frac{4F_{act}\sigma}{24}$, fig.~\ref{potential_fig}.c inset $x_{f}=0, 1$). 

We evaluate the validity of this approach via the mode center-to-center distance. We found that the histogram for data simulated using equation~\ref{netAlpha} consistently displays a mode center-to-center distance in agreement with the intended diameter (fig~\ref{potential_fig}.a, c). Additionally, the data is sharply distributed around the intended diameter as is shown through the cumulative distribution function (CDF), which increases steeply proximal to the intended diameter. We found that our implementation for the repulsive force can consistently be considered hard-sphere, irrespective of activity or particle fraction (see fig.~\ref{potential_fig}a, c, teal diamonds).

Implementing the interparticle potential according to 
equation~\ref{netAlpha} affects the maximum timestep possible for a
given simulation ($\tau_{LJ}=\frac{\sigma^{2}\gamma}{\epsilon}$ and
$dt_{max}=10^{-5}\tau_{LJ}$). Since we chose to vary the swim 
velocity in our model, we also need to vary the potential well 
depth and the timestep. Therefore, we present our results using the
temperature-dependent (and invariant) timescale for the persistence
of motion, $\tau_{r}$.
We also performed convergence studies on our model to determine the
necessary ratio of $l_{p}:l_{box}$ to avoid finite size effects at 
the densities and activities used in this study (see ESI, fig. S1).

\subsection{Simulation Parameters and definitions} \label{simpar}

Simulations were performed using the GPU-enabled Molecular Dynamics package available in HOOMD-blue~\cite{Anderson2008, Glaser2015}. We initialized particles randomly (allowing a slight particle overlap, $\delta=0.70$) and then equilibrated the system via Brownian dynamics for $10\tau_{B}$. After equilibration, simulations were run for $100\tau_{B}$. Total system area fraction was constant at $\phi=0.6$ (where $\phi=\phi_{f}+\phi_{s}$). At this area fraction, systems were above the minimum density required to display MIPS ($\phi_{crit.}\approx0.45$ for monodisperse active systems).

The onset of MIPS was first analyzed by a cluster algorithm~\cite{freud} with a cutoff that was calibrated on simulations of monodisperse active particles ($Pe=30$) below the critical activity for MIPS ($Pe_{Critical}\approx45$). At this activity, the system has not undergone MIPS, however, small, short-lived clusters will regularly form and fall apart. Thus it gives us the largest signal for transient clusters that are not truly phase separated both in terms of size and lifetime.
For simulations proximal to the binodal, our algorithm was often insufficient in delineating which systems were phase-separated due to the large fluctuations in the number and size of transient clusters. In this regime, in addition to the algorithm, we examined the phase behavior by visual inspection. Phase-boundary lines are indicated clearly in our plots. Additionally, all simulation images and videos were generated using OVITO~\cite{Stukowski2010VisualizationTool}.

Our approach in this paper is to introduce the idea of a continuum for active matter systems, with limiting cases defined by the activity of the less active (or \lq slow\rq) species. When the less active particles are Brownian we have an active/passive mixture. As the activity of the \lq slow\rq \ species increases we obtain a mixture of two types of particles where each species has a distinct activity and velocities fast and slow. Finally as the activity of each species becomes the same, we get a monodisperse active system. Monodisperse active and active/passive systems can be viewed as a subset of this larger active/active continuum. 
Here, we move between the two extreme cases via two parameters: the fast particle fraction ($x_f$) and the slow particle activity ($Pe_s$). 
In our investigation of the active/active phase space we have found two additional parameters which help in examining the trends of active/active systems. First, the \lq net activity\rq \ which is simply an average of each particle's activity weighted by its particle fraction:
\begin{equation}\label{eqnNetAct}
    Pe_{net}\equiv x_{s}Pe_{s}+x_{f}Pe_{f}.
\end{equation}
Note that as either of the examined parameters ($x_{f}$ or $Pe_{s}$) is increased the net activity also increases
However, $Pe_{net}$ does not account for the relative activity between the two species.  
To this end we also use the activity ratio, which is a ratio of the slow particle activity to the fast:
\begin{equation}\label{eqnActRat}
    Pe_{R}\equiv \frac{Pe_{s}}{Pe_{f}}
\end{equation}
and varies between zero (an active/passive mixture) and one (a monodisperse active system).

\begin{figure*}
    \includegraphics[width=1\textwidth]{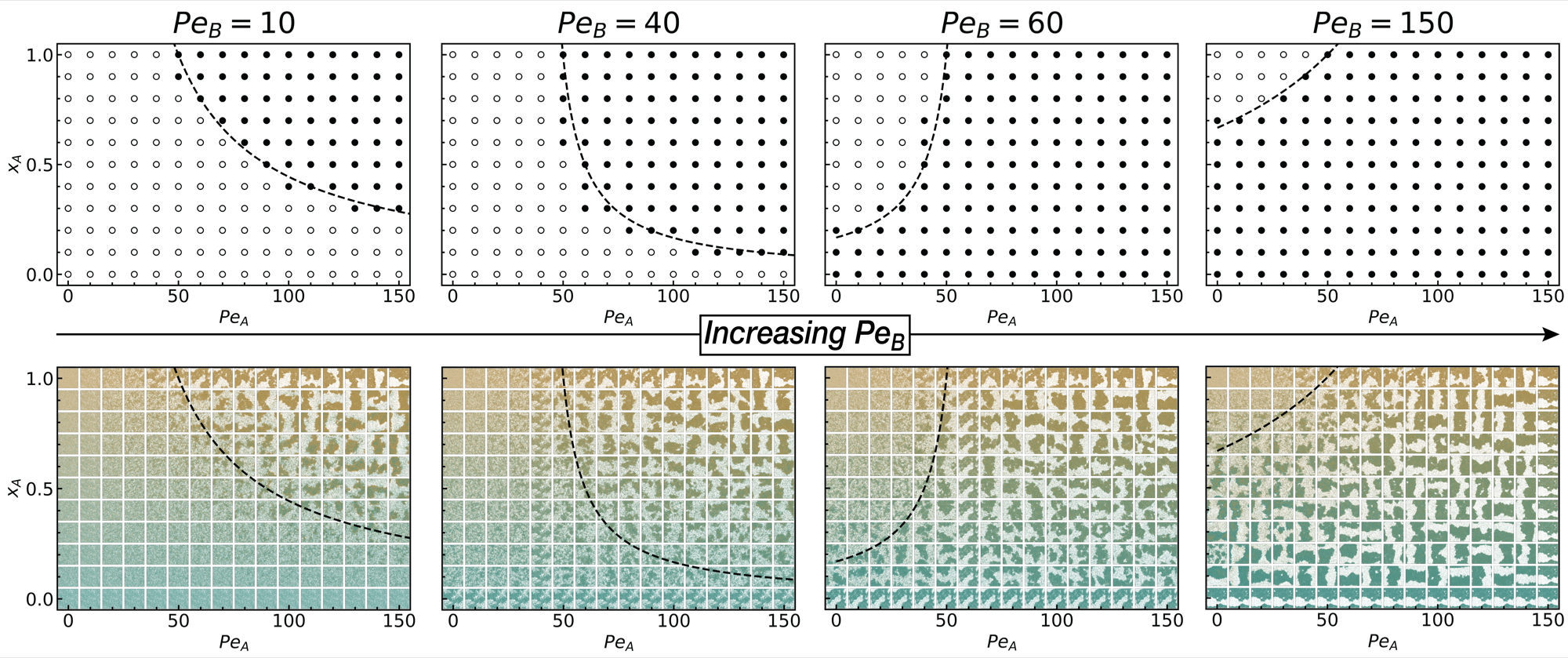}
    \caption{
    Top row: Phase diagram in the $x_A$-$Pe_A$ plane for $Pe_B=10, 40, 60, 150$. Filled symbols denote phase-separated systems, as determined by our MIPS-identifying algorithm, and open symbols denote a gaseous steady-state. The dashed lines indicate phase boundaries predicted by the active/active equation for $\phi_0$ in table 1, which is the theoretical binodal according to our kinetic theory. There is good agreement between the simulations and theory. Bottom row: the same phase diagrams are shown illustrated by simulation (final) snapshots after 100$\tau_B$, where `A' particles are shown in gold and `B' particles in teal. (See ESI, Fig. S2 for the full parameter range).
    }\label{phasePlane}
\end{figure*}

\section{Background}\label{background}

The onset of MIPS was first examined in monodisperse active systems using the ABP model~\cite{Redner2013a}. In their study, Redner \textit{et al.}~\cite{Redner2013a} calculated the densities of the dense and dilute phases versus activity at various total system area fractions. At all area fractions, increasing activity corresponded to more particles in the dense phase (and, commensurately, fewer particles in the gas phase). 

Expanding the phase space, Stenhammar et al. examined binary mixtures of active and passive particles~\cite{Stenhammar2015}. Aside from the total area fraction and particle activity, an additional parameter was introduced, the particle fraction of each species ($x_{act}+x_{pass}=1$). It was shown that by increasing either the magnitude of the active species' activity, or the fraction of active particles, the system became more likely to phase separate and MIPS was achieved at lower area fractions of active particles. It was also observed that the distribution of particles \textit{within} the dense phase was not homogeneous; instead, there were small domains of predominantly active or predominantly passive particles. The domains were such that the interior of the cluster consisted mostly of trapped passive particles while the edges consisted mostly of active particles. For MIPS to persist, the body axis of particles at the cluster edge must form an acute angle with respect to a normal to the cluster surface (or else they simply leave the cluster)~\cite{Redner2013a, Redner2013, Patch2018}. As a passive particle does not exert an active force it cannot maintain the cluster edge.

Both studies developed a kinetic theory, to predict the critical area fraction for MIPS based upon the parameters of activity and concentration, thus producing nonequilibrium phase diagrams. 
In both cases, an assumption was made that there is a steady-state equilibrium between an infinite, hexagonally close-packed, dense phase and a gas phase. Additionally, for an active/passive mixture, the density of active particles in the gas phase and in the dense phase were assumed to be equally proportional to the active particle fraction ($\phi_{g}x_{act}=\phi_{act,g}$). For a monodisperse active system at steady-state, the derived rates of particle adsorption (from the gas to the dense phase) and desorption (from dense to gas phase) are equal (Table 1). Thus, by writing expressions for these terms and equating them, Redner et al obtained a theoretical description for the density of each phase~\cite{Redner2013a}.

%\vspace*{-2mm}
\renewcommand\arraystretch{2.25}
\[
\begin{array}{c | ccc}\label{kineticTheories}
 & \text{Active\cite{Redner2013a}} & \text{Active/passive\cite{Stenhammar2015}} & \text{Active/active}\\
 \hline
k_{\text{in}}: &  
{\displaystyle\frac{4\phi_{g}v_{0}}{\pi^{2}\sigma^{2}}} &
{\displaystyle\frac{4\phi_{A,g}v_{0}}{\pi^{2}\sigma^{2}}} &
{\displaystyle\frac{4(\phi_{s,g}v_{s} + \phi_{f,g}v_{f})}{\pi^{2}\sigma^{2}}}\\
k_{\text{out}}: &
{\displaystyle\frac{\kappa D_{r}}{\sigma}} &
{\displaystyle\frac{\kappa D_{r}}{\sigma}} &
{\displaystyle\frac{\kappa D_{r}}{\sigma}}\\
\phi_{0}: &
{\displaystyle\frac{3\pi^{2}\kappa}{4Pe}} &
{\displaystyle\frac{3\pi^{2}\kappa}{4x_{A}Pe}} &
{\displaystyle\frac{3\pi^{2}\kappa}{4(x_{s}Pe_{s}+(1-x_{s})Pe_{f})}}
\end{array}
\]
%\vspace*{-5mm}
%\begingroup
%\caption{\noindent Table 1. Expressions for rates of adsorption on to and desorption from the dense phase as well as the critical area fraction for MIPS, for: monodisperse active, active/passive, and active/active systems. The more complex active/passive and active/active expressions reduce to the simpler monodisperse case.}
%\endgroup

\noindent Table 1. Expressions for rates of adsorption on to and desorption from the dense phase as well as the critical area fraction for MIPS, for: monodisperse active, active/passive, and active/active systems. The more complex active/passive and active/active expressions reduce to the simpler monodisperse case.

\vspace{5mm}

\noindent In introducing a second (passive) species, these rates remain equal, but the new expression included the effect of the particle fraction of each species~\cite{Stenhammar2015}. Additionally, for both studies, particle desorption was assumed to occur in cascading, \lq avalanche\rq \ events, resulting in a fitting parameter $\kappa$. It is important to note that this approach assumes the existence of an HCP dense phase, and as such, it is not valid at very high or very low area fractions. 

\section{Results} \label{results}

\subsection{Kinetic theory \& phase diagram} \label{phasetheory}

By removing the restriction that the slow particles be passive, we introduced another parameter, the activity of the second species. To populate this new phase diagram we restricted our study to total area fraction $\phi=0.6$ to ensure that we were above the critical area fraction for monodisperse MIPS ($\approx$0.45). We simulated ${N=15,000}$ particles and systematically varied three parameters: the activity of each species A, B, $(Pe_{A,B}=[0,150])$ and the particle fraction $(x_{A}=[0.0, 1.0])$, resulting in 1,232 simulations in total (fig.~\ref{phasePlane}). Figure~\ref{phasePlane} shows four representative slices of the three dimensional phase diagram for this space, where $Pe_{B}$ was held constant at 10, 40, 60 and 150. Each $x_{A}$-$Pe_{A}$ plane is a phase diagram showing whether MIPS occurred, as computed by our MIPS algorithm and denoted by filled points. In the bottom row of figure~\ref{phasePlane}, the same phase diagrams are shown illustrated by simulation (final) snapshots after 100$\tau_B$.  On any given plane, increasing the fraction of the more active species (moving up along a column where $Pe_{A}>Pe_{B}$ or down a column where $Pe_{A}<Pe_{B}$), or increasing the activity of species A (to the right on a given row) gives parameters more conducive to MIPS. Similarly, as we increased the activity of species B (fig.~\ref{phasePlane}, left to right), more of the phase plane becomes phase separated. To distill these data, we found that the propensity for a system to undergo MIPS increased as the activity of any constituent species increased or as the fraction of the more active species increased (fig.~\ref{phasePlane}).

We also extended the existing kinetic theories for monodisperse active~\cite{Redner2013a} and passive/active~\cite{Stenhammar2015} systems, summarized in section~\ref{background} to include a second active species. In examining the rate of adsorption ($k_{in}$) and desorption ($k_{out}$), we continued to make the assumption of previous models that desorption from the dense phase to the gas is governed exclusively by rotational diffusion (and is therefore identical for each particle, given they are the same size and equilibrated in the same heat bath). However, adsorption from the gas onto the dense phase depends on particle activity, as the more active particles have a larger persistence length $(l_{p})$ and therefore collide with the dense phase more readily. This expansion results in the expressions for $k_{in}$ and $k_{out}$ for active/active systems that are presented in table 1; $\phi_{g}$ is the density of a given species in the gas phase (subscript g), and $v$, the intrinsic swim-speed of a given particle~\cite{Redner2013a}.

\begin{figure*}[ht!]
    \includegraphics[width=0.95\textwidth,trim={0 0 0 0},clip]
    {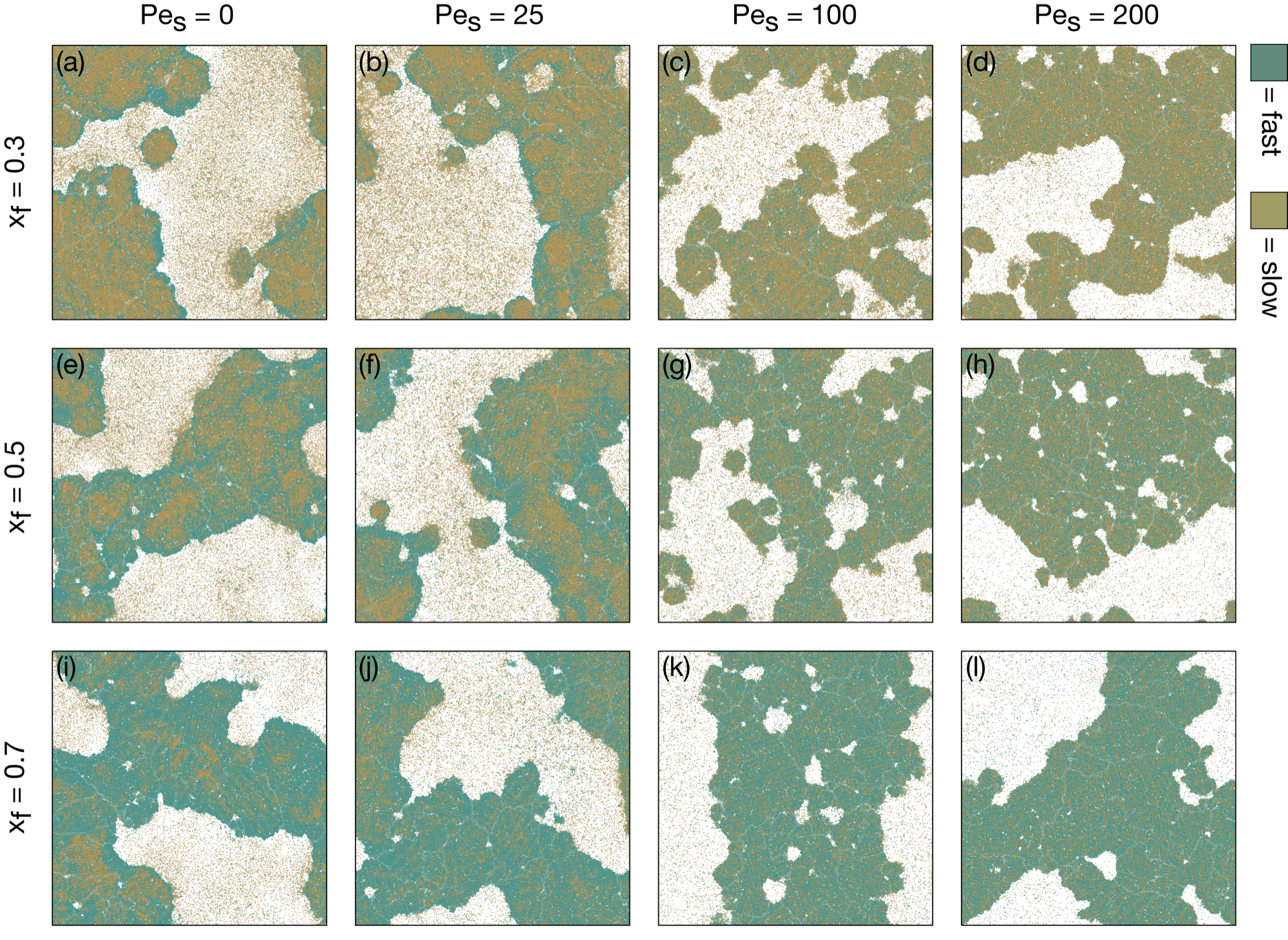}
    \caption{
    (\href{https://drive.google.com/file/d/1A9boYIIiFx_9gI2bNh-jaYVCMO6s7cJQ/view?usp=sharing}{Video online}) Simulations of active Brownian particles with distinct slow (gold) and fast ($Pe_{f}=500$, teal) particle fractions and activities. All snapshots are taken after steady-state had been reached, $\tau_{B}=100$. The particle fraction is constant in each row: $x_{f}=0.3$, $0.5$, and $0.7$; illustrating majority slow, equal and majority fast mixtures respectively. The activity of the slow species is constant in each column: $Pe_{s}=0$, $25$, $100$, and $200$ which is representative of the observed emergent behaviors in this space. The activity of the fast species is fixed at $Pe_f=500$.
    }\label{qualPhaseSeg}
\end{figure*}

We then considered dimensionless variables and extended what has been reported as a \lq binodal relation\rq \ ($\phi_{act,g}=x_{act}\phi_{0}$) developed by Stenhammar \textit{et al.}~\cite{Stenhammar2015} to include a second active species, which gives the binodal conditions: $\phi_{A,g}=x_{A}\phi_{0}$ and likewise $\phi_{B,g}=(1-x_{A})\phi_{0}$ for active species A and B. The two together can be summarized by the sole relation initially developed by Redner \textit{et al.}~\cite{Redner2013a}, $\phi_{0}=\phi_{A,g}+\phi_{B,g}=\phi_{g}$. 
The relation produced by equating the flux at the cluster edge does, in fact, predict very reliably, the onset of MIPS in simulation. However, we should note that this is not a binodal in the strict thermodynamic sense, namely there is no equilibration of a thermodynamic quantity between phases. 
Substituting these relations in we can obtain an expression that relates activity, particle fraction and area fraction to phase separation (table 1), where, if $\phi_{0} \le \phi$ the system will undergo MIPS. This theoretical \lq binodal\rq \ should ideally separate closed (phase separated) points from open points, as identified from our simulation data. Our extension of this kinetic theory demonstrates strong agreement with simulation ($>99\%$) as seen in figure~\ref{phasePlane}, where, the theory is represented by the dashed line on each plane.  Note that we used $\kappa=4.05$, the same value used by ref.~\citenum{Stenhammar2015} which produced the best fit across the phase space examined in this study.

\subsection{Phase behavior} \label{tuned}

In figure~\ref{qualPhaseSeg} we show characteristic snapshots of the system at different slow particle activity and particle fraction. As the systems are highly dynamic, we suggest the reader watch the videos for each snapshot as well (ESI). These snapshots qualitatively show how the emergent phenomena characteristic of active/passive mixtures transitions to the distinct steady-state behavior of monodisperse active systems. The columns in figure~\ref{qualPhaseSeg} correspond to four different slow activities that are representative of this phase space: $Pe_{s}=0$ (Brownian), $25$ ($<Pe_{crit}$ below critical activity for MIPS if the system were monodisperse), $100$ ($>Pe_{crit}$ above critical activity for MIPS if the system were monodisperse), and $200$ (about half of the fast species activity). 
Rows correspond to three particle fractions, $x_{f}=0.3$, $0.5$, and $0.7$ in order to examine majority slow, equal, and majority fast mixtures respectively.

In the case of an active/passive mixture (fig.~\ref{qualPhaseSeg}, left column (a),(e),(i)) we found behavior in agreement with previous studies~\cite{Stenhammar2015}, both in terms of structure and of dynamics, an additional verification of our velocity-variant model.

We now discuss the effect of changing the particle fraction for an active/passive system. 
For a system composed entirely of passive particles ($x_{f}=0$), MIPS did not occur and the system remained a homogeneous gas undergoing random Brownian diffusion. If we substitute a small fraction of passive particles for active particles (\textit{e.g.} $x_{f}=0.05$) the resulting system still remained in the gaseous state. The small number of fast active intruders left temporary wakes behind them as they swam through the majority passive gas (ESI, \href{https://drive.google.com/file/d/1VdvvWw0Fxdx2_fgWghgrJzXeEoGJCED1/view?usp=sharing}{movie} and fig. S4). Increasing the fraction of fast particles further ($x_{f}=0.1$), small clusters formed and were quickly annihilated as a thin outer edge of fast particles pushed through the passive interior (ESI, \href{https://drive.google.com/file/d/176070JKbBg3233luemvMhk36889JsZv_/view?usp=sharing}{movie} and fig. S5).
At the critical particle fraction $(x_{f}=0.15)$ phase separation occurred; there was still a majority of passive species and a small number of very fast active particles. This was perhaps the most volatile system to undergo sustained MIPS in that clusters engaged in fission and fusion events repeatedly, yet, the dense phase was never absent entirely and the system still coarsened in time. Fast particles predominated at the cluster edge and pushed the slower particles into the dense phase (similar to fig.~\ref{qualPhaseSeg}(a)). The clusters here seemed to be qualitatively different to the clusters that occur through MIPS in typical monodisperse active systems~\cite{Redner2013a} due to their volatility. Increasing the fraction of fast particles further $(x_{f}>0.15)$ did not prevent partial segregation by particle type, but, the system behavior gradually moved toward a monodisperse (fast) active system becoming markedly less volatile as fast particle fraction increased (fig.~\ref{qualPhaseSeg}(e), (i)).

By turning on the activity of the slow species (fig.~\ref{qualPhaseSeg}, (b)), we observed how an active slow component altered the system behavior. Note that a monodisperse system prepared at $Pe_{s}=25$ with the same total system density is below critical activity for MIPS. We found the same trends with respect to particle fraction as in the active/passive case. Below critical fast particle fraction (now $x_{f,crit}=0.05$), we found a gas with few active tracers where the path left by the fast component now collapsed more quickly (as the slow component did not rely on Brownian diffusion to consume the void left by the fast particle). As fast particle fraction increased (fig.3 (f), (j)), we obtained the same behaviors present in the active/passive case with a seeming decrease in system volatility. The partial segregation by particle type observed in active/passive systems, persists when the slow species is active, but to a lesser extent. Ultimately, there is no discontinuous jump in system behavior with respect to slow activity and particle fraction. 

As we continued increasing the slow activity (for both $Pe_{s}=100$, $200$, see fig. 3 (c),(d)), the system underwent MIPS regardless of fast particle fraction. Additionally, the distribution of each particle type appears homogeneous, with no prevalent species at the cluster edge nor in domains in the cluster interior. For systems prepared at slow activity above the critical activity, we found that the main function of particle fraction was in tuning the extent of phase separation between the compositions of a monodisperse slow system (for $x_{f}\rightarrow 0$) and a monodisperse fast system (for $x_{f}\rightarrow 1$). As the slow particle activity approached that of the fast species, system volatility decreased.

As one of the characteristics of a mixture is the distribution of particles, with the most extreme case being a partially segregated active/passive mixture, then we can qualitatively distinguish two regimes. One where phase separation also shows segregation by type (active/passive is included here) and a second regime where there is a more homogeneous dense phase (which, notably occurs outside of monodisperse active systems, $Pe_{s}\neq Pe_{f}$). In other words, at low slow activity the system behaves like a passive/active mixture and at high slow activity the system behaves like a monodisperse active system. 

\subsection{Steady state composition of dense phase}\label{steady}

\begin{figure*}
    \includegraphics[width=1.0\textwidth,trim={0 0 0 0},clip]
    {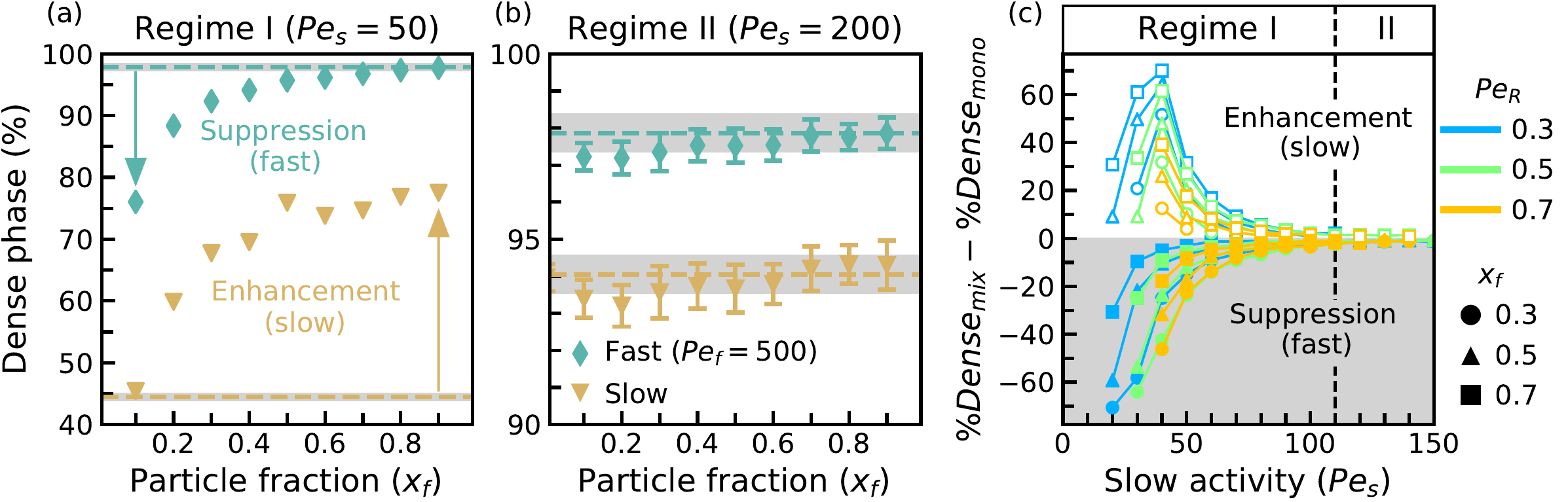}
    \caption{
    Steady-state participation in the dense phase for both fast and slow species. (a, b) Percentage of each particle type in the dense phase for different particle fraction in (a) regime I $(Pe_{s}=50)$ and (b) regime II $(Pe_{s}=200)$ compared to a monodisperse system of each particle type (dashed lines). Error bars are shown when larger than the symbol size and in grey shading for monodisperse systems.  (a) In  regime I slow species experience enhancement and fast species exhibit suppression with respect to a monodisperse system of the same total area fraction. (b) In regime II fast and slow species assume the dense phase participation (within error) of a monodisperse fast or slow system respectively. These data are condensed in (c) a plot of the difference in the percent participation in the dense phase between a mixed and monodisperse system at distinct activity ratios and particle fractions. Negative values (filled symbols) indicate suppression of the fast component and positive values (unfilled symbols) show enhancement of the slow component. Regime I persists for systems that undergo MIPS with slow activity up to $Pe_{s}\approx 110$ and activity ratio less than unity. Beyond this point, each species assumes the steady-state participation in the dense phase correspondent to a monodisperse simulation at the same activity and total system area fraction.
    }\label{mixingRegimes}
\end{figure*}

To gain more insight and obtain a quantitative understanding of the two regimes observed, we computed the dense-phase composition after steady-state had been reached. To examine the effect of mixing, we compared the composition of the equivalent monodisperse system for each simulation, \textit{i.e.} one with the same total area fraction and activity. For example, in figure~\ref{mixingRegimes}a, diamond teal symbols show the percentage of fast particles in the dense phase as a function of the fast particle fraction, for constant slow activity $Pe_s=50$ and constant fast activity $Pe_f=500$. The teal dashed line at $\approx 98\%$ shows the percentage of fast particles in the dense phase for a monodisperse active system of fast particles at $Pe=500$. Similarly, for slow particles, gold triangles show the percentage of slow particles in the dense phase as a function of the fast particle fraction, for constant slow activity $Pe_s=50$ and constant fast activity $Pe_f=500$. The brown dashed line at $\approx 45\%$ shows the percentage of slow particles in the dense phase for a monodisperse active system of slow particles with $Pe=50$. 

In regime I $(Pe_{s}\lesssim 110)$, we observed deviations in the steady-state composition of each species in the dense phase with respect to a monodisperse system (of the same activity, fig.~\ref{mixingRegimes}a). 
We found that the faster species experienced a suppression in the percentage of particles participating in the dense phase relative to its monodisperse counterpart. The slower species experienced the opposite effect, the steady-state participation in the dense phase was enhanced upon being mixed with a faster species. 

In regime II $(Pe_{s} \gtrsim 110)$, the enhancement and suppression characteristic of regime I disappeared entirely. Instead, each species, regardless of what it was being mixed with, assumed the same percentage of particles participating in the dense phase as a monodisperse system of the same activity (fig.~\ref{mixingRegimes} b). This result is counterintuitive, above a specific slow activity, each species behaves as if it weren't mixed at all. The significance of the slow particle activity at which this regime change occurs, as well as the dependence of this value on parameters not examined in this study (\textit{i.e.} the total area fraction) are of great interest to the authors and require further study.

Our data gives a clear relationship for which parameters affect the magnitude of the enhancement or suppression effect that is found in regime I. The trends in these data are most evident in figure~\ref{mixingRegimes}c, where we show percent difference between the dense phase of a mixture and a monodisperse system, versus the slow particle activity. The fast species is always suppressed (relative to a monodisperse fast system, bottom half of fig.~\ref{mixingRegimes}c) just as a slow species can only be enhanced (relative to a monodisperse slow system, top half of fig.~\ref{mixingRegimes}c). We see that for any set slow activity, the extent of the enhancement and/or suppression was dependent on both the particle fraction and activity ratio. Moreover, the particle fraction was found to control the dominant behavior (suppression/enhancement) for a given system.
A system composed of a majority of fast particles predominantly undergoes enhancement of the slow species (fig.~\ref{mixingRegimes}(a), $x_{f}=0.9$) and vice versa for a predominantly slow system (fig.~\ref{mixingRegimes}(a), $x_{f}=0.1$). The activity ratio was found to control the magnitude of either behavior. For low activity ratio ($Pe_{R}\approx 0$), both enhancement and suppression are amplified (fig.~\ref{mixingRegimes}a, $Pe_{R}=0.1$) and for activity ratio close to unity the effect is greatly reduced (and evidently is non-existent at $Pe_{R}=1$). Additionally, both behaviors need not occur in a given system at the same time (\textit{e.g.} in fig.~\ref{mixingRegimes}(a), at $x_{f}\ge 0.5$ the slow species is being enhanced while the fast species is not being suppressed).

To contextualize this result we relate to ABP/PBP systems. In these systems the slow species is being \lq pushed\rq \ into the dense phase by the fast species (enhanced relative to a monodisperse passive system which does not undergo MIPS at all). We have shown that this finding is not exclusive to an ABP/PBP mixture, in fact, this effect can be found at slow particle activity far above the critical activity for a monodisperse system ($Pe_{crit}\approx45$ for $\phi=0.6$). 
The extent of the enhancement/suppression decays asymptotically with respect to the slow activity. As such, our \lq regime II\rq \ simply indicates that the asymptotic decay has approached close enough to zero so that the effect of enhancement or suppression is within standard deviation of the steady-state cluster participation.

\begin{figure}[ht!]
    \includegraphics[width=0.5\textwidth,trim={0 0 0 0},clip]
    {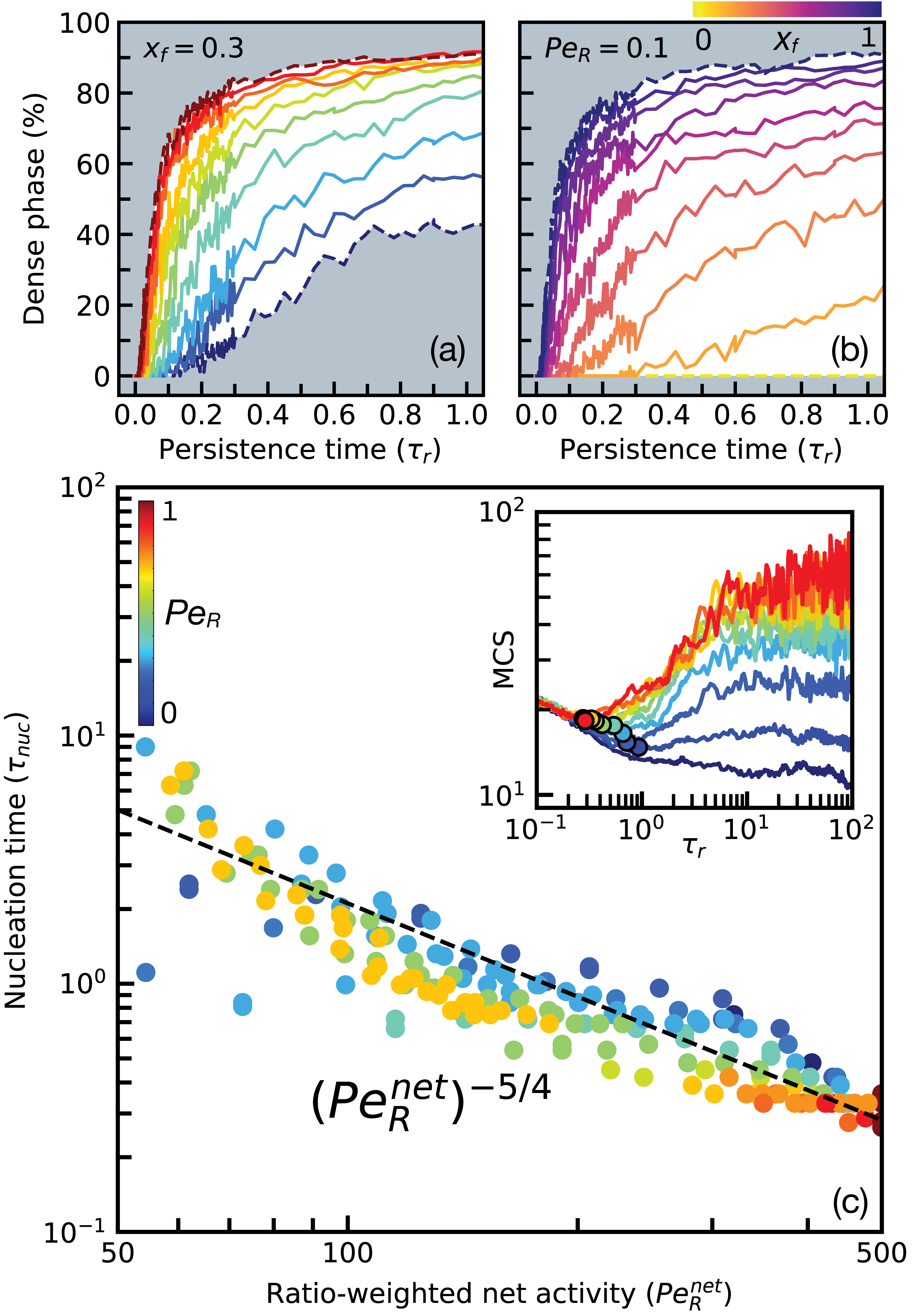}
    \caption{
        Time-resolved participation of all particles in the dense phase at constant particle fraction (a, $x_{f}=03$) showing the effect of changing slow particle activity. Colorbar indicates the activity ratio of each simulation ($Pe_{f}=500$). Constant activity ratio (b, $Pe_{R}=0.1$) illustrates the effect of varying the particle fraction (where shade indicates each simulated particle fraction). Dynamic data shows a distinct relationship between nucleation time and net activity (c). Data is collapsed from a wide variety of activity ratios ($Pe_{R}$), particle fractions ($x_{f}$), and fast particle activities ($Pe_{f}$). Inset shows computation of the nucleation time, taken as the local minimum in the mean cluster size vs. time.
    }\label{time_collapse}
\end{figure}
\subsection{Dynamic behavior}\label{dynamic}
We then examined the formation and growth of the dense phase on short timescales. To do this, we measured the number of particles in the dense phase as a percentage over the total number of particles. We found that both fast and slow species nucleated the dense phase at the same time, however, the rate of growth of each species was distinct where the more active component would add to the dense phase more rapidly (ESI, fig. S6). Additionally, we report that the early-stage dense phase composition did not track that of a monodisperse system of either the slow or fast particle type. 
This effect is observed upon varying activity ratio at constant particle fraction and for varying particle fraction at constant activity ratio (given that $x_{f}$ and $Pe_{R}\neq 0, 1$). At constant particle fraction (fig.~\ref{time_collapse}(a)), increasing the activity ratio moves dynamic percent of particles in the dense phase closer to a monodisperse system of fast particles (maroon dashed line). The same behavior was observed at constant activity ratio (fig.~\ref{time_collapse}(b)) for increasing fast particle fraction (dark purple dashed line).

Both of these observations can be summarised via the net activity, where, by increasing the net activity, a system either moves a homogeneous gas closer to the binodal line or causes a greater percentage of particles to join the dense phase, at a quicker rate. While active/active mixtures at steady state have been shown to take on the characteristics of an active/passive mixture or monodisperse active system, the behavior of binary active systems dynamically, is distinct from either extreme case.

As demonstrated in figure~\ref{time_collapse}(c), the net activity also affects the nucleation time of an active system. We extracted the time that the mean cluster size, for each simulation, began to increase (fig.~\ref{time_collapse}c, inset) and plotted this against each system's net activity on a log-log scale (fig.~\ref{time_collapse}c). The initial nucleation event has a power law dependence on the net activity (fig.~\ref{time_collapse}c, exponent ~$\frac{5}{4}$), where a higher net activity corresponds to a faster nucleation time. So, the net activity (an intrinsic quantity) sets a single time for cluster nucleation, independent of individual species activity. Growth of the cluster thereafter is controlled by each species' activity, where an increase in the net activity corresponds to more rapid coarsening and growth.

\subsection{Net activity}\label{netActivity}

\begin{figure*}
    \includegraphics[width=1.0\textwidth,trim={0 0 0 0},clip]
    {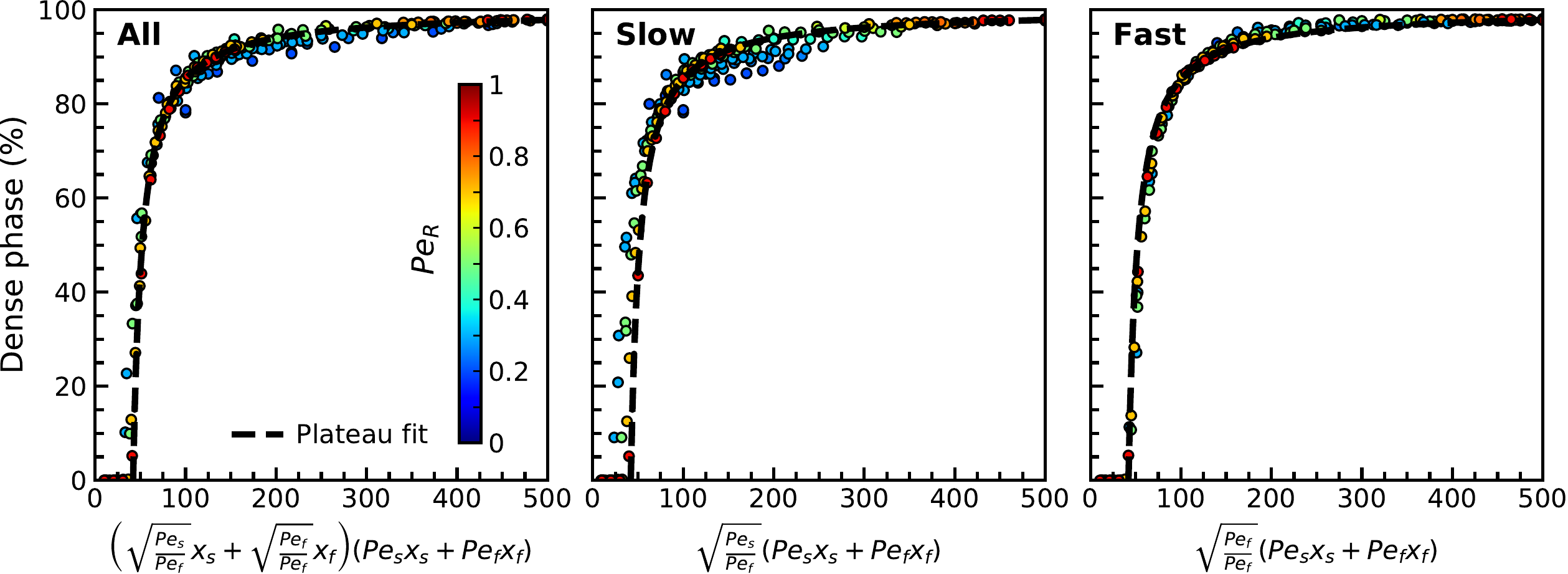}
    \caption{
    Steady-state percentage of particles in the dense phase for all simulations performed in this study, plotted against a weighted average of the constituent particle activities. The data collapses onto a plateau function $(ax/b+x)$ where $a=100$ (maximum percent in the dense phase) and $b=10$.
    }\label{netActivityCollapse}
\end{figure*}

To this point, we have observed the importance of the net activity in a number of ways: setting the repulsive potential, as a component of our kinetic theory, and as it provides a clear representation of the trends in the space of active/active mixtures (\textit{e.g.} a high net activity corresponds to faster nucleation times). We now show that the net activity also has predictive utility. In plotting the net activity versus the percent of fast particles in the dense phase (for all simulations with $N \ge 10^{5}$) we found a reasonable collapse of all data onto a single curve (ESI, fig. S7). However, the data for the dense phase participation of slow particles exhibited an additional dependence on the activity ratio (ESI, fig. S7).  We were able to further collapse this data by multiplying a weighted average of the magnitude of activities by a modified weighted average of the activity ratios (fig.~\ref{netActivityCollapse}). 
\begin{equation}\label{weightCollapse}
    Pe_{R}^{net}=\left(x_{s}\sqrt{\frac{Pe_{s}}{Pe_{f}}}+x_{f}\sqrt{\frac{Pe_{f}}{Pe_{f}}}\right)\big(x_{s}Pe_{s}+x_{f}Pe_{f}\big)
\end{equation}
We then fit the data from figure~\ref{netActivityCollapse} to a plateau function, and obtained a relationship,
\begin{equation}\label{platFunc}
    f(x_{f}, Pe_{s}, Pe_{f}) = \frac{100 \ Pe_{R}^{net}}{10 + Pe_{R}^{net}},
\end{equation}
which can compute the steady-state percent of each particle type which participates in the dense phase without the need for simulation.

\section{Conclusion}\label{conclusion}

We computationally studied mixtures of active particles of different activities and thus velocities (fast and slow). Our method maintained the intended particle diameter for hard spheres via a parameter-dependent repulsive force. We calculated the binodal envelope for MIPS in the particle fraction-activity phase space and extended existing kinetic theories on monodisperse active and active/passive systems to incorporate a second active species. Our extended kinetic theory was in good agreement with simulations for the binary systems studied here. 

In analyzing the dynamics and steady-state behaviors of active/active systems we summarize the main findings and discuss implications. 
Emergent phenomena known from monodisperse active matter and from active/passive mixtures (\textit{e.g.} MIPS, segregation by particle type, volatility \textit{etc.}) can be tuned to appear or vanish in active/active mixtures showing that these phenomena are part of a continuum of behaviors. We can however categorize into two mixing regimes: (i)Regime I: the faster/slower particles experience suppressed/enhanced participation in the dense phase due to mixing.
(ii) Regime II: each particle type participates in the dense phase as if it were a monodisperse system. The two regimes and the transition between them indicates that there is a tolerance of the emergent behaviors with respect to the two species' activities, i.e. there is a whole range of parameters for which we see each regime.  
It is interesting to note, for example, that the transition between the two regimes for the systems studied here occurs when $Pe_s\approx 110$ regardless of the fast activity. In other words, the two activities in a mixture can be quite different in magnitude and still the system would act like a monodisperse active system. This could explain the robustness of natural swarms where we might expect that differences in the velocities of animals/organisms would not immediately result in a collapse of the group behavior. 
Moreover, this transition between a regime where there is segregation by type due to distinct activities and a regime where the system acts like a monodisperse homogeneous one could have implications in biological systems such as liquid-liquid phase separation in the cell~\cite{llps}. For example, it could be that distinct activities contribute (together with other interactions) to phase separation and a change in the activities of two species could regulate transitioning from a gas (no MIPS) to a phase separated state with segregated domains (regime I) to a mixed dense phase (regime II).  

Furthermore, two quantities naturally emerge from our analysis, capturing the physics of these mixtures: the net activity (eq.~\ref{eqnNetAct}) and a `ratio-weighted net activity' (eq.~\ref{weightCollapse}). The net activity gives the repulsive strength and is a natural result of a simple kinetic theory for this system. The slightly more complex `ratio-weighted net activity' predicts, a priori, the participation of each species in the dense phase, as well as the timescale for nucleation in active/passive, active/active, and monodisperse active systems. 

While our results demonstrate the complex behavior that is accessible to multi-component active mixtures, the work presented here is only the first step. A number of interesting future directions emerge.
We expect that a greater degree of control is achievable in synthetic active matter via the introduction of additional active species (as is the case in biological contexts). Furthermore, a mixture of active species (distinct in their activity) gives an intriguing setup for the examination of potential thermodynamic quantities that set the characteristics of active matter, such as the active pressure and its implications regarding an equation of state\cite{Takatori2016e,Fily2018,Solon2015f,Solon2015g}. 

\section*{Acknowledgments}
\noindent This material is based upon work supported by the National Science Foundation Graduate Research Fellowship under Grant No. (NSF grant number DGE-1650116). DK and TK are thankful for conversations with Ehssan Nazockdast, Julien Tailleur and John Brady. We also thank Thomas Dombrowski, Ian Seim and Clayton Casper for insightful comments. 
% \balance

% \clearpage

%%%REFERENCES%%%
\bibliography{manual_refs} % this is copy and pasted from the mendeley import

\end{document}